\documentclass[]{revtex4}

\usepackage[dvips]{graphicx}

\begin{document}


\newcommand{\va}{\vspace{5mm}}
\newcommand{\vb}{\vspace{10mm}}
\newcommand{\vs}{\vspace{15mm}}


\newcommand{\be}{\begin{equation}}
\newcommand{\ee}{\end{equation}}
\newcommand{\ba}{\begin{eqnarray}}
\newcommand{\ea}{\end{eqnarray}}
\newcommand{\NL}{\nonumber \\}

\newcommand{\binom}[2]{ \left( \begin{array}{c} #1 \\ #2 \end{array} \right)}
\newcommand{\E}[1]{ \langle #1 \rangle }
\newcommand{\ov}[1]{ \overline{#1} }



\title{How low is the thermodynamical limit?}


\author{Tam${\rm \acute{a}}$s  S. Bir${\rm \acute{o}}$, G${\rm \acute{a}}$bor 
Purcsel}
\affiliation{KFKI Research Institute for Particle and Nuclear Physics\\
H-1525 Budapest, Pf 49, Hungary}
\email{tsbiro@sunserv.kfki.hu}
\author{Berndt M${\rm \ddot{u}}$ller}
\affiliation{Physics Department, Duke University \\
Durham, NC 27708, USA}


\begin{abstract}
We analyze how can some dynamical process lead to exponential
(Boltzmannian) distribution without instantaneous thermal
equilibrium with a heat bath. We present a model for parton
dressing which re-combines the exponential from cut power law
minijet distributions fitted to pQCD as a limiting distribution.
Thermal models and power law tails are interpreted within the 
framework of the more general Tsallis distribution.
\end{abstract}


\maketitle


\section{Introduction}
Not only in heavy ion collisions, but also in $pp$ and even in
$e^+e^-$ collisions exponential transverse mass spectra has
been observed besides a power law tail at relatively high
transverse momenta. The exponential shape at medium $p_T$
was often interpreted as an indication that the source of
the observed hadrons can be described by a temperature related
to the slope of these spectra. This view is most prominent
in the thermal models\cite{THERMAL} which assume a thermally equilibrated
state, well described by the familiar equilibrium
thermodynamics.

The experimental data from CERN SPS and RHIC show an exponential
shape in the transverse mass spectrum
best fitted by  $\exp((m-m_T)/T)$, where
$m_T=\sqrt{m^2+p_T^2}$ is the transverse mass of a particle
with mass $m$\cite{EXP}. The stiffness of these spectra, $T$, has still
to be corrected for collective flow effects: for high momenta
a relativistic blue-shift factor, for low momenta a mass
dependent kinetic flow energy addition occurs in this parameter.
The conjectured temperature, $T_0$, of a thermally equilibrated state
is therefore usually higher than the directly observed slope
parameter, $T$. Characteristic values at RHIC are $T \approx 140$ MeV
and $T_0 \approx 210$ MeV, so the assumed temperature is well above
the color deconfinement temperature predicted by recent lattice
QCD calculations.

There is, however, doubt whether a thermally equilibrated state can be
formed in such a short time, during which the final state fireball at
these high bombarding energies develops. There is no known hadronic
process which would lead to a thermal state from a far off-equilibrium
initial situation on a time scale of $1-2$ fm/c. It is strongly tempting
to consider quark level mechanisms as reason for the observed exponential
spectra. Several such mechanisms has already been suggested on a subhadronic
level. Ref.\cite{Bialas} considered the Schwinger mechanism with a Gaussian
fluctuating string tension, which leads to exponential $m_T$ spectra.
Having color ropes in mind, the effective string tension may have a
Gaussian distribution due to a random color chargation process in the limit
of high color. 
Another idea is to consider the fluctuation of
temperature, distributed according to a Gamma distribution\cite{Wilk}. In this
case the average of exponentials leads to a Tsallis distribution, which is
still well approximated by an exponential at soft momenta, while ends in a
power-law tail at high momenta. 
In this paper we shall have a closer look
on a recently suggested hadronization mechanism, on the parton 
recombination\cite{RECOMB},
as a possible source of the experimentally observed exponential spectra.
In this case we begin with cut power-law minijet distributions\cite{PARTON}, 
having a form of the canonical Tsallis distribution\cite{TSALIS}, 
and combine their limiting
distribution for many-fold recombination\cite{PLB}.

According to this goal 
first we review how the Gaussian distribution emerges as a limiting
distribution and recall the central limit theorem of statistics.
Then we consider a simple example not falling under the conditions of
this renown theorem, but still leading to a non-trivial limiting
distribution. Finally we relate this result to parton recombination and to
the canonical Tsallis distribution, which gives an improved
quantitative description of experimental findings compared to the naive thermal
model. 

\va
\section{The central limit theorem}

\begin{figure}
\vspace*{-10mm}
\begin{center}
\includegraphics[width=0.25\textwidth]{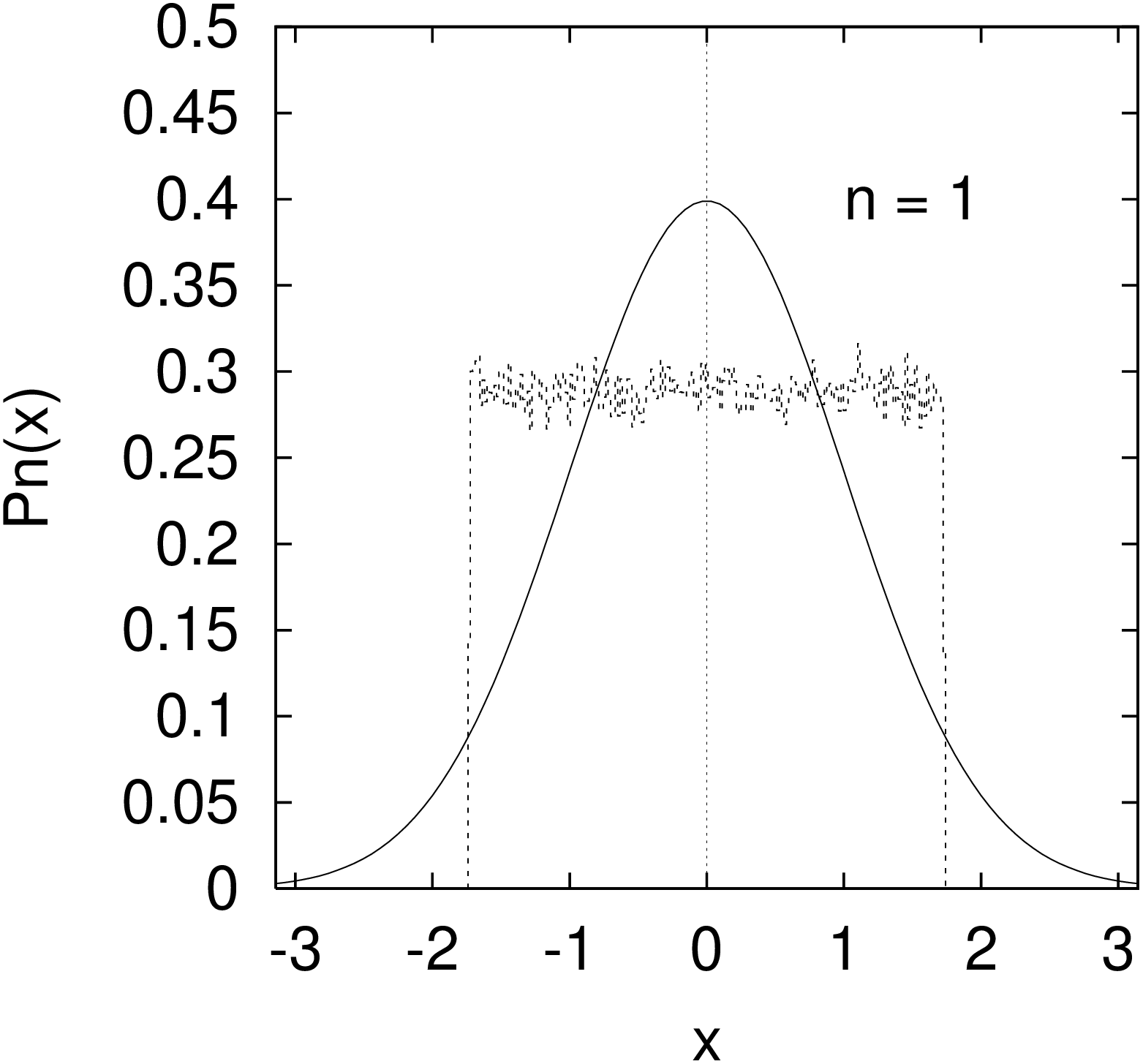}\includegraphics[width=0.25\textwidth]{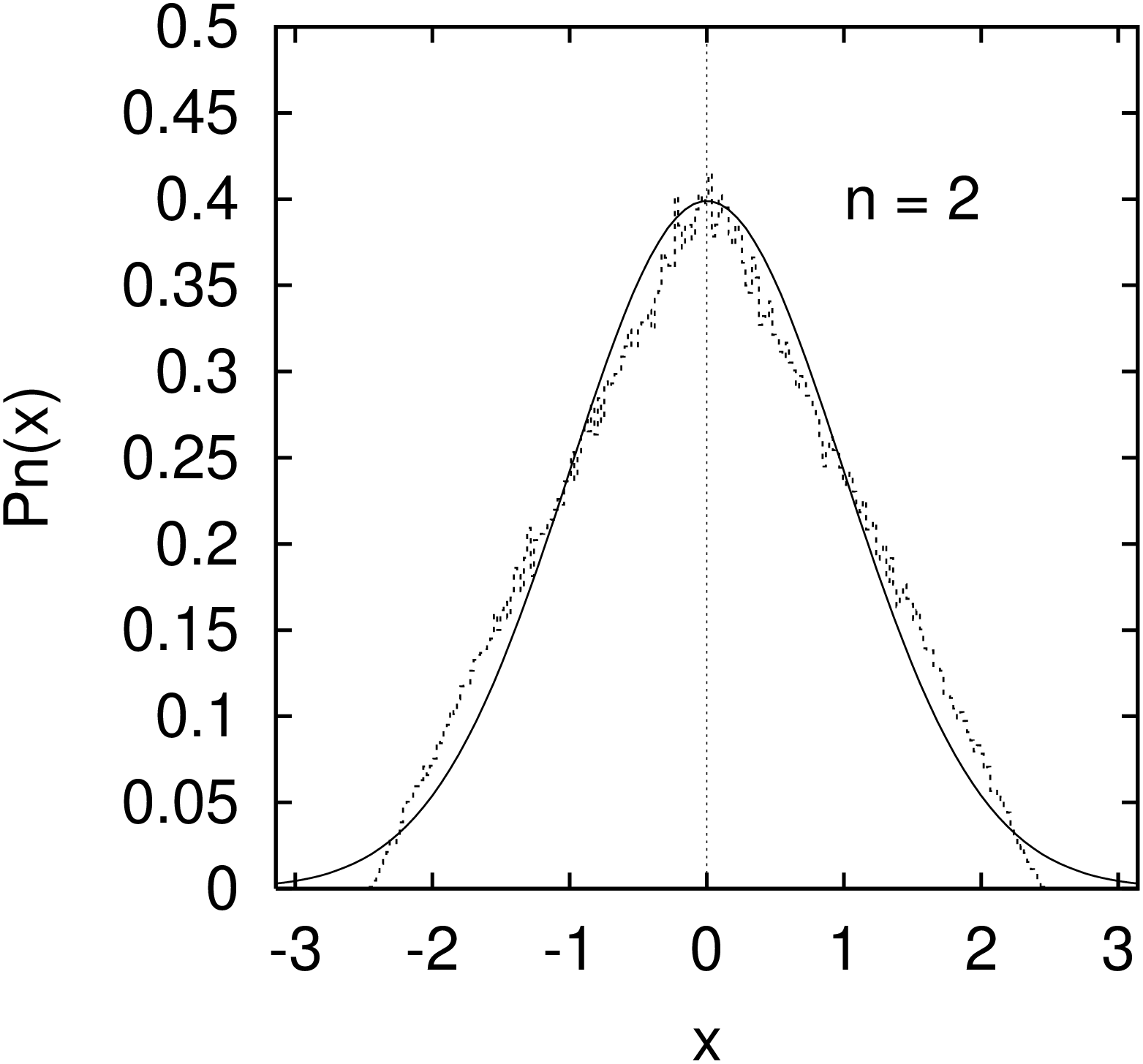}
\includegraphics[width=0.25\textwidth]{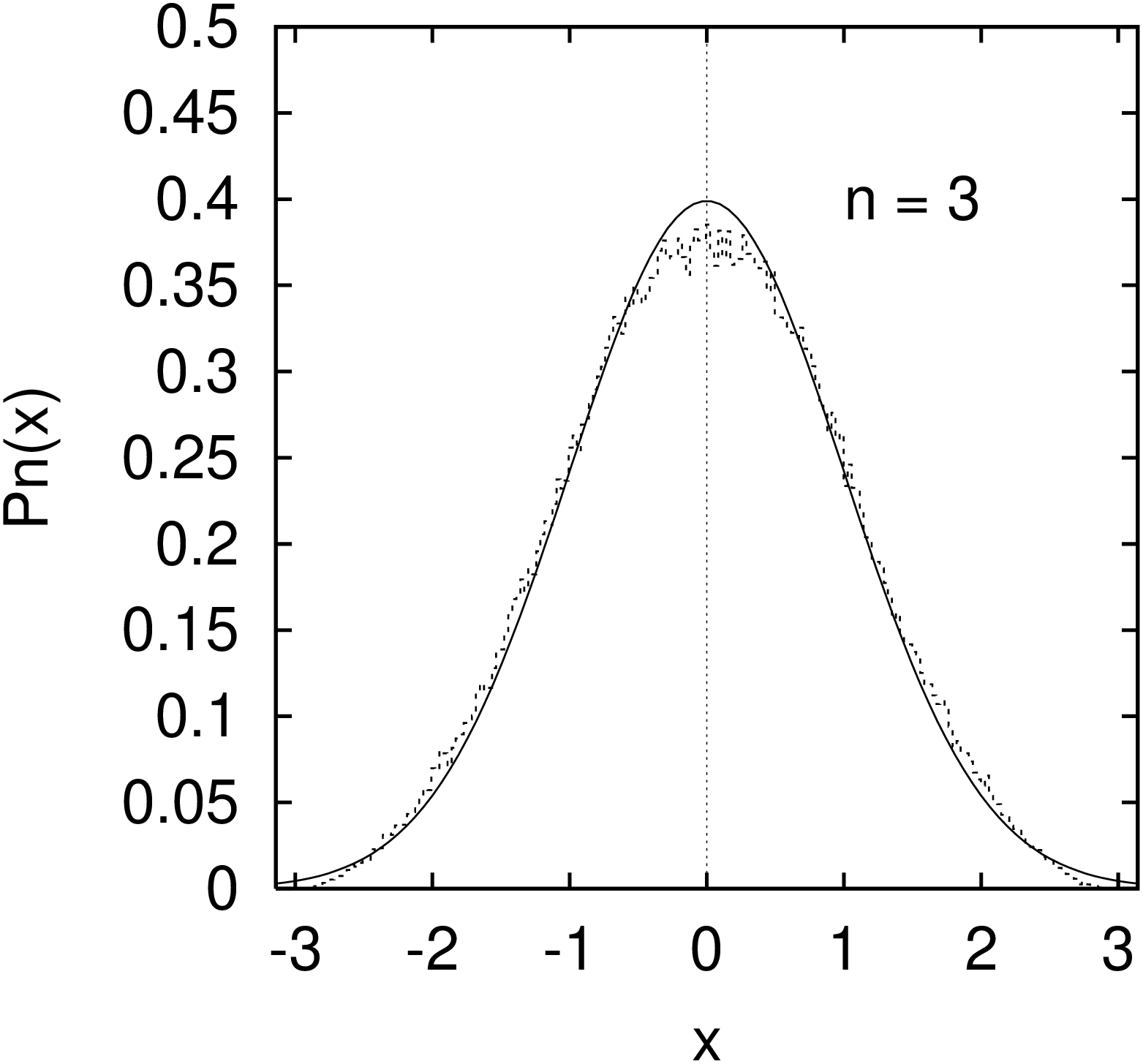}\includegraphics[width=0.25\textwidth]{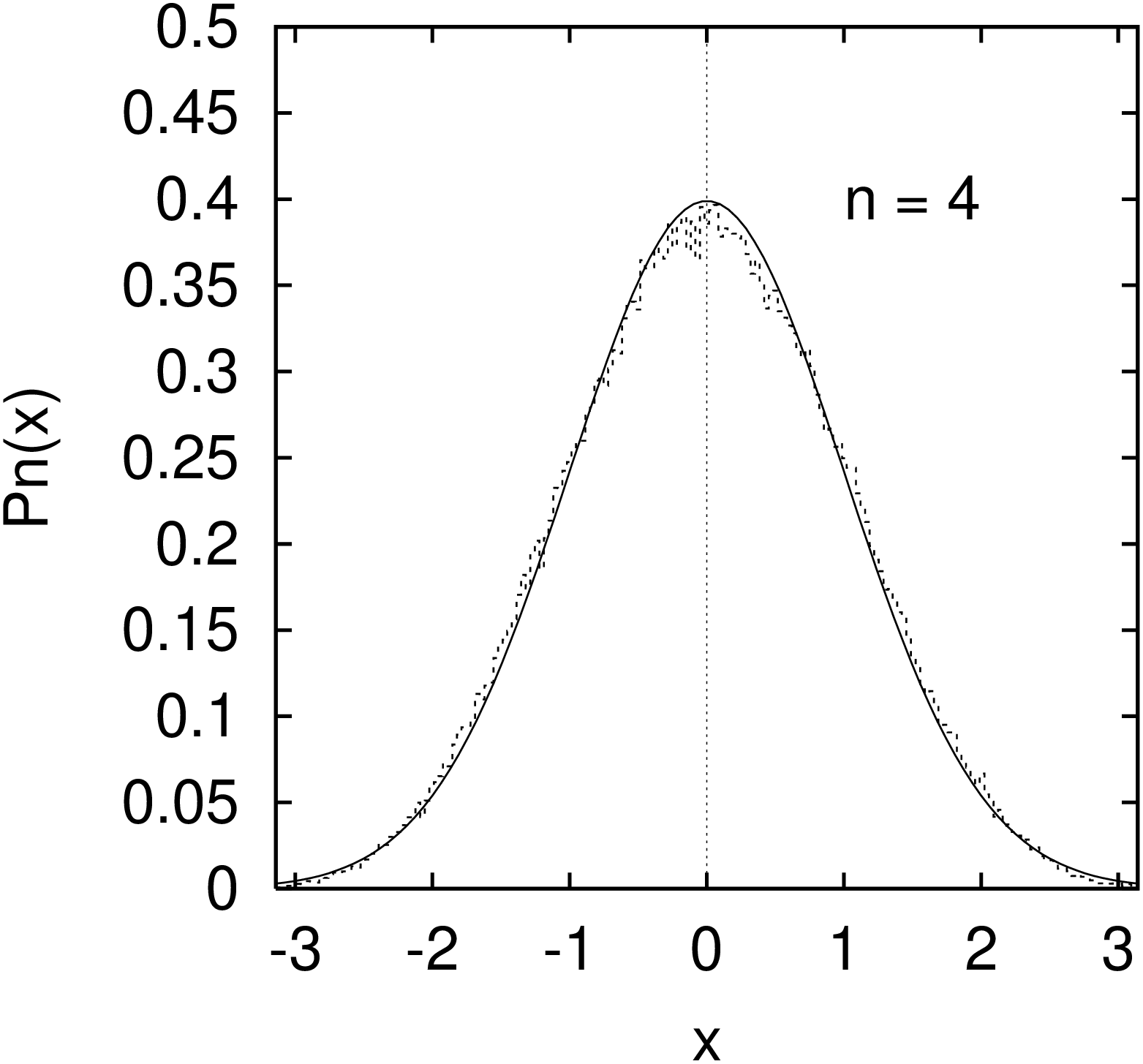}
\end{center}
\vspace{-2mm}
\caption{\label{Fig1} 
 Comparison of histograms of $m = 200 000$ sums of $n$ uniform
 random deviates in $(-1,1)$ scaled with $\sqrt{3/n}$ and the
 limiting Gauss distribution $\frac{1}{\sqrt{2\pi}}\exp(-x^2/2)$.
}
\end{figure}

In order to demonstrate, how fast a Gaussian limiting distribution may
develop, let us first consider the distribution of the scaled sum
of uniform random deviates.
Let $x_i$ be a uniform random deviate in  $(-1,1)$
and {$P_n(x)$} the distribution of
$  x \; = \;  \sqrt{\frac{3}{n}} \: \sum_{i=1}^{n} \, x_i$.
The statement of the cited theorem is that in the limit of infinite $n$
the distribution of the variable $x$ exactly becomes the normal
distribution.
\be
  P_{\infty}(x) \; = \; \frac{1}{\sqrt{2\pi}} \; e^{-x^2/2}.
\ee
Before we review how and why this happens it is educating to make some
guesses how large $n$ actually should be in order to mistake such a
distribution for a Gaussian. The answer is surprisingly low (cf. Fig.\ref{Fig1}).
The $n=1$ case is the uniform distribution, the $n=2$ has a triangular shape.
The finiteness of the sample ($m=200000$) smoothes the differences out, it is
therefore not easy to distinguish already the $n=3$ or $n=4$ case from the
Gaussian in a numerical experiment.


Now we turn to the proof of the limiting distribution being a Gaussian.
Consider the Fourier transform of $P_n(x)$:
\be
 \tilde{P}_n(k) \: = \: \int_{-\infty}^{\infty}\!\!\! dx \, e^{ikx}
 \, \prod_{i=1}^n \left( \frac{1}{2} \int_{-1}^{1}\!\! dx_i \right) \, 
 \delta(x-\sqrt{\frac{3}{n}}\sum_{j=1}^n x_j).
\ee
After x-integration it turns to be a product of $n$ equal factors:
\be
 \tilde{P}_n(k) \; = \; \left( \frac{\sin (k \sqrt{3/n})}{k\sqrt{3/n}} \right)^n.
\ee
The large $n$ limit of this expression is a Gaussian:
\be
 \tilde{P}_{\infty}(k) \; = \; \lim_{n \rightarrow \infty}
 \left( 1 - \frac{k^2}{2n} + \ldots \right)^n \; \longrightarrow \;
 e^{-k^2/2}.
\ee
Fourier transforming back to x-space gives the required statement.
The phenomenon is more general. Let $x$ be a weighted sum of $n$
independent random variables:
\be
 P_n(x) = \int (\prod_{i=1}^n w(x_i) dx_i) \; \delta(x - a_n \sum_{j=1}^n x_j). 
\ee
each distributed according to the same probability density $w(x)$.
Then the so called central moments, which describe correlations, 
\be
c_j(n) = \left. \left( \frac{\partial}{i\partial k} \right)^j \log \tilde{P}_n(k) \right|_{k=0},
\ee
can be obtained easily from the central moments $c_j(1)$ of the $P_1(x)=w(x)$
distributions using $\log \tilde{P}_n(k) = n \log \tilde{w}(ka_n)$:
$  c_j(n) = n a_n^j c_j(1)$.   
Whenever $a_n\propto 1/\sqrt{n}$, 
$c_j(n) \propto n^{1-j/2}c_j(1) \longrightarrow 0 \: {\rm as} \: 
     n \rightarrow \infty$ 
for $j > 2$.
This completes the proff of the Central Limit Theorem.


There are exemptions to this theorem. Whenever
$\log \tilde{P}_{\infty}(k)$ is not differentiable arbirary times
in the vicinity of $k=0$, or not the proper scaling factor $a_n$ is used, or
the base distribution $w(x)$ has a long tail not approaching zero
fast enough.  In the last case there can be a limiting distribution, 
it is just not Gaussian.
As an interesting example we study the convolution of normalized Lorentzians
\be
 w(x) \: = \: \frac{R}{\pi}\: \frac{1}{x^2+R^2}
\ee
The logarithm of the Fourier transform is not differentiable at $k=0$:
$ \log \tilde{w}(k) = -|k|R$.
The n-fold convolution's Fourier transform also has a non-differentiable 
logarithm, $ \log \tilde{P}_n(k) \: = \: - nR |k| a_n$.
It can, however, be found a finite limiting distribution for
($n \rightarrow \infty$) if {$a_n = 1/n $}.
In this case the Lorentzian is a fixed point of the convolution and scaling.
The limiting Lorentzian distribution is also long-tailed, so
the central moments diverge.
Aiming at physical applications we have to regularize. 
We use a  small parameter, $m$, for this purpose
and at the end take the nontrivial double limit
{$m \rightarrow 0$ and $n \rightarrow \infty.$}
Starting with regularized Lorentzians, the Fourier transform has the
form of the experimentally observed $m-m_T$ exponential,
\be
 \tilde{w}(k) = \exp(\frac{m  -\sqrt{k^2+m^2}}{T} )
\ee
with $T = 1/R$.
Now $\log \tilde{w}(k) = R(m - \sqrt{k^2+m^2})$ with vanishing odd
central moments, and with regular even central moments,
$ c_{2j}(1) \propto \frac{R}{m^{2j-1}}$.
The limiting distribution is obtained by keeping {$ M = nm $} constant
and scaling with $a_n=1/n$ again. It has the following Fourier spectrum
\be 
\tilde{P}_{\infty}(k) \: = \: e^{(M-\sqrt{k^2+M^2})/T}.
\ee
This resembles the original regularized distribution, but with an
eventually finite mass $M$. The low and high momentum limits are
Gaussian and pure exponential, respectively.
%
%

\va
\section{Parton dressing to exponential spectra}

Combining hadrons from many partons also results in a limiting
distribution, which is not Gaussian.
We consider an additive quark model where the hadron mass
is made of $n$ approximatevly equal partons: {$ M = \sum E_i = n m.$}
The variable,
$ x = \frac{\sum E_i x_i}{\sum E_i} = \frac{1}{n} \sum x_i$
is the center of energy coordinate. Its limiting distribution,
$P_{\infty}(x)$ can be viewed as the hadron wave function squared.
Its Fourier transform, $\tilde{P}(k)$ is proportional to the hadron spectrum.

While constructing from "equal-right" partons, we have
$ \tilde{P}_n(k) = \tilde{w}(k/n)^n$. The particular cases $n=2$ and $n=3$
of this formula have been used in parton recombination models of
hadronization recently.
The partons before hadronization show a cut power law
transverse momentum distribution, as fitted to minijets obtained in
the framework of pQCD based calculations:
\be
 \tilde{w}(k) = a \left( 1 + \sqrt{k^2+m^2}/b \right)^{-c} 
\ee
Here $b$ and $c$ are parameters of the parton distribution,
in no way connected to a thermal state, $m$ is the transverse 
mass at zero rapidity, and finally $a$ can be obtained from
the normalization condition, $\tilde{w}(0)=1$.
Note that this distribution is a {power law} for large transverse
momenta
$ \lim_{|k| \gg m, b} \: \tilde{w}(k) \: = \: ab^c \, |k|^{-c}$.

We construct a constituent quark and eventually hadrons
from $n$ such partons, keeping
the total mass $M=nm$ fixed. We arrive at
\be
 \tilde{P}_n(k) = a^n \left(1 + \frac{1}{nb}\sqrt{k^2+M^2} \right)^{-nc} 
\ee
In the $n \rightarrow \infty$ limit this gives the exponential
distribution in the transverse mass
\be
 \tilde{P}_{\infty}(k) \: = \: 
 	\exp \left( \frac{M - \sqrt{k^2+M^2}}{b/c} \right). 
\ee
This is exponential of the reduced transverse mass
$M_T-M=\sqrt{k^2+M^2}-M$ has a slope $T = b/c$.
This is not a temperature, it is not related to any energy
exchange with a heat reservoir. The large parameter by which one
approaches the limiting distribution is the foldness of the
recombination; in the present model the number of independent
primordial partons dressing to a hadron constituent.

\va
\section{{Tsallis distribution and thermal model}}

The cut power law used in minijet fits and modified as a function
of energy in the hadronization model presented above accidentally
coincides with the canonical Tsallis distribution.
This distribution can be derived from a generalization of the
Boltzmann entropy formula first suggested by Tsallis:
\be
S_q \: = \: \frac{1}{1-q} \sum_i (w_i^q-w_i ) 
\ee
In order to relate to our model we use $q=1-1/n$. We get
\be
 S_q = n \sum_i w_i (w_i^{-1/n}-1) \longrightarrow - \sum_i w_i \ln w_i. 
\ee
The canonical distribution maximizes the entropy with
the following constraints: $\sum_i w_i=1$ and $\sum_i w_i \varepsilon_i = E$.
The use of a lagrange multipliers $\beta$ leads to the following
distribution which minimizes $S_q-\beta E$:
\be
 w_i = \frac{1}{Z} \left( 1 + \frac{\beta\varepsilon_i}{n} \right)^{-n} 
\ee
with $Z = \sum_i (1+\beta\varepsilon_i/n)^{-n}$ for the proper normalization.

Based on this distribution the free energy can be derived as well as
relations reminding to rhe familiar thermodynamical relations.
The Tsallis distribution can hence be regarded as a generalization
of equilibrium thermodynamics. Among its physical realizations
it is particularly interesting the case of finite heat reservoirs.

In view of this formalism we interpret the 
one parton distribution as a canonical Tsallis distribution,
\be
w_1 = (1 + \frac{E}{b})^{-c} = w_{{\rm Tsallis}}(E;T,1-1/c) 
\ee
with $T=b/c$.
Recombination of $n$ such partons leads to another Tsallis
distribution with the same ''temperature'', but closer to the
equilibrium case $q=1$:
\be
 w_n = w_1(E/n)^n = (1 + \frac{E}{nb})^{-nc} =
	w_{{\rm Tsallis}}(E;T,1-1/(nc)). 
\ee
Arriving at the finally observed hadrons with $c\approx 8$ 
and $n=2 \ldots 6$ one gets really close to the limiting exponential
\be
w_{\infty} = w_{{\rm Tsallis}}(E;T,1)= e^{-E/T}. 
\ee
This property may explain the relative, and so far
unexplained, success of the thermal model.

Finally it is interesting to explore the potential of the Tsallis
distribution with respect to particle numbers and the energy per
particle, since these are the most profound predictions of the
thermal model. For the sake of simplicity we consider here the
analytically solvable case of massless particles.

The number of a particle in a volume $V$ with internal degeneracy
factor $d$ is given by the integral of the canonical Tsallis
distribution,
\be
 N = \frac{d}{2\pi^2} V \, \int_0^{\infty} p^2dp \: 
 \left(1 + \frac{E}{b} \right)^{-c}.
\ee
In the massless case, $E = |p|$, and we get
\be
 N = \frac{d}{2\pi^2} V b^3 \,  \frac{2}{(c-1)(c-2)(c-3)}.
\ee
In general this is a bigger number for finite $c$ than for
the Boltzmann distribution at $nc \rightarrow \infty$:
$N(c)/N(\infty)  > 1$.
This ratio is $2.4, 1.5, 1.3$ for $c = 8, 16, 24.$

The energy is given by another integral,
\be
 E = \frac{d}{2\pi^2} V \, \int_0^{\infty} p^2dp \: E \,
 \left(1 + \frac{E}{b} \right)^{-c}
\ee
which can as well be calculated in a closed form for massless particles,
$E = |p|$:
\be
 E = \frac{d}{2\pi^2} V b^4 \,  \frac{6}{(c-1)(c-2)(c-3)(c-4)}
\ee
The energy per particle becomes
$ E/N = \frac{3b}{c-4} $ resulting in
 {$E/N \approx 1.07$ GeV} for $b=1.39$ GeV and $c=7.9$, and
 at the same time predicting a slope $T \approx 176$ MeV
 for the $m_T$ exponential at low momenta $p_T \ll bc \approx 10$ GeV.


\va
\section { Conclusion}

We reviewed limiting distributions of familiar short tailed
and exceptional long tailed distributions. We have seen, that
already a few, $n = 3-4$ independent random comopnents may lead
to a distribution which is hard to distinguish from the limiting
case on finite, fluctuating data samples.

Non Gaussian limiting distributions exist with altered (non $1/\sqrt{n}$) 
scaling laws. For the high energy hadronization process it is
particularly interesting to consider cut power law distributions,
which in the limiting case combine to exponential spectra.
This happens with the $1/n$ scaling, pointing out a possible
interpretation as hadron spectra showing the Fourier transform of
the distribution of the center of energy of the recombining partons.

Minijet partons obtained using pQCD processes, and actually also
experimentally observed spectra in $pp$ collisions are 
Tsallis distributed at $T \approx 170  - 200$ MeV, 
and with $E/N \approx 1$ GeV.  These two numbers can be simultanously
described by a Tsallis distribution with the parameters
$b=1.39$ GeV and $c=7.9$, when considering massless particles.
The same simple consideration fails for the familiar Boltzmann
distribution of the thermal equilibrium with a factor of $2$.
At finite mass these integrals can be calculated only numerically.
Considering massive particles with a mass around the pion mass,
our results do not change dramatically. At the mass of 1 GeV,
however, the Boltzmann distribution reaches $E/N \approx 1$ GeV
and the Tsallis distribution overshoots with a factor of 2.


\begin{acknowledgments}
This work has been supported by the Goethe University, Frankfurt and
by the Hungarian National Research Fund OTKA (T034269). Special thanks
to Professor Walter Greiner for organizing this NATO Advanced Study
Institute meeting at Kemer, Turkey, which location created a perfectly 
relaxing and inspiring atmosphere.
\end{acknowledgments}




\begin{thebibliography}{1}


\bibitem{THERMAL}
	J.Letessier, J.Rafelski, A.Tounsi, Phys.Lett.B {\bf 328}, 499, (1999);
	J.Cleymans, H.Oeschler, K.Redlich, Phys.Lett.B {\bf 485},
	27, (2000);
	K.Redlich, K.Werner, Phys.Rev.Lett. {\bf 88}, 202501, (2001);
	W.Bro\-ni\-ow\-ski, A.Baran, W.Florkowski, nucl-th/0212053, nucl-th/0212052;
	P.Braun-M\"unziger, K.Redlich, J.Stachel,
	in Quark Gluon Plasma 3, eds.: R.Hwa, X-N.Wang, World Scientific
	Publ.Co., (2003), nucl-th/0304013;
	F.Becattini, G.Pettini, Nucl.Phys.A {\bf 715}, 557, (2003).

\bibitem{EXP} PHENIX collaboration, nucl-ex/0307022, nucl-ex/0307010,
	nucl-ex/0304022, nucl-ex/0212014, Acta Phys.Hung. Heavy Ion Physics
	{\bf 15}, 291, (2002);
	STAR collaboration, hep-ex/0306056, nucl-ex/0306024;
 	NA49 collaboration, nucl-ex/0306022.

\bibitem{Bialas} A.Bialas, Phys.Lett.B {\bf 466}, 301, (1999).

\bibitem{Wilk} G.Wilk, Z.Wlodarczyk, Phys.Rev.Lett. {\bf 84}, 2770, (2000);
	hep-ph/0002145; hep-ph/0004250.

\bibitem{RECOMB} R.J.Fries, B.M\"uller, C.Nonaka, S.A.Bass,
	nucl-th/0301087, nucl-th/0305079, nucl-th/0306027.

\bibitem{PARTON} R.J.Fries, B.M\"uller, D.K.Srivastava, Phys.Rev.Lett.
	{\bf 90}, 132301, (2003);
	D.K.Srivastava, C.Gale, R.J.Fries, Phys.Rev.C {\bf 67}, 034903, (2003).

\bibitem{TSALIS} C.Tsallis, J.Stat.Phys. {\bf 52}, 479, (1988),
	Physica A {\bf 221}, 277, (1995); A.Plastino, A.R.Plastino, 
	Brazilian J. Phys. {\bf 29}, 50, (1999).

\bibitem{PLB} T.S.Biro, B.M\"uller, hep-ph/0309052.

\end{thebibliography}
\end{document}